\documentclass{acmsiggraph}

\usepackage[utf8]{inputenc}
\usepackage{overpic}
\usepackage{amsmath}
\usepackage[usenames,dvipsnames]{color} 
\usepackage{soul}
\usepackage[normalem]{ulem}
\usepackage{hyperref}
\usepackage{xfrac} 

\newcommand{\mypara}[1]{\noindent{\textbf{#1.}}}
\newcommand{\vnudge}{\vspace{-.2in}}

\definecolor{userColor1}{rgb}{0.05,0.85,0.22}
\definecolor{userColor2}{rgb}{0.67,0.25,0.80}
\definecolor{userColor3}{rgb}{0.13,0.15,0.84}
\definecolor{hiddenColor}{rgb}{0.7,0.7,0.7}
\definecolor{redColor}{rgb}{1,0,0}
\definecolor{outlineColor}{rgb}{0.65,0.27,0.80}
\definecolor{orangeColor}{rgb}{1,0.54,0}
\definecolor{gray}{rgb}{0.7,0.7,0.7}

\newcommand{\holger}[1]{ \textcolor{userColor1}{\textbf{Holger:} {#1}}}
\newcommand{\mira}[1]{ \textcolor{userColor2}{\textbf{Mira:} {#1}}}
\newcommand{\james}[1]{ \textcolor{userColor3}{\textbf{James:} {#1}}}

\long\def\ignorethis#1{}
\long\def\hide#1{\begin{center}{\textbf{\textcolor{hiddenColor}{... original Text hidden...}}}\end{center}}

\title{How2Sketch: Generating Easy-To-Follow Tutorials for Sketching 3D Objects}

\author{ \\ University College London, Han, Holger, Mira, Niloy}
\pdfauthor{}

\author{ James W. Hennessey$^1$ \and
Han Liu$^{2}$  \and
Holger Winnemöller$^3$  \and
Mira Dontcheva$^3$  \and
Niloy J. Mitra$^1$ \\ \\\hspace{-15cm}\text{$^{1}$University College London \hspace{0.75cm}$^{2}$KAUST \hspace{0.75cm}
$^{3}$Adobe Research }}

\newcommand{\setcand}[1]{\mathcal{C}_{*#1}}

\begin{document}
\setcounter{page}{1}
 \teaser{
   \includegraphics[width=\textwidth]{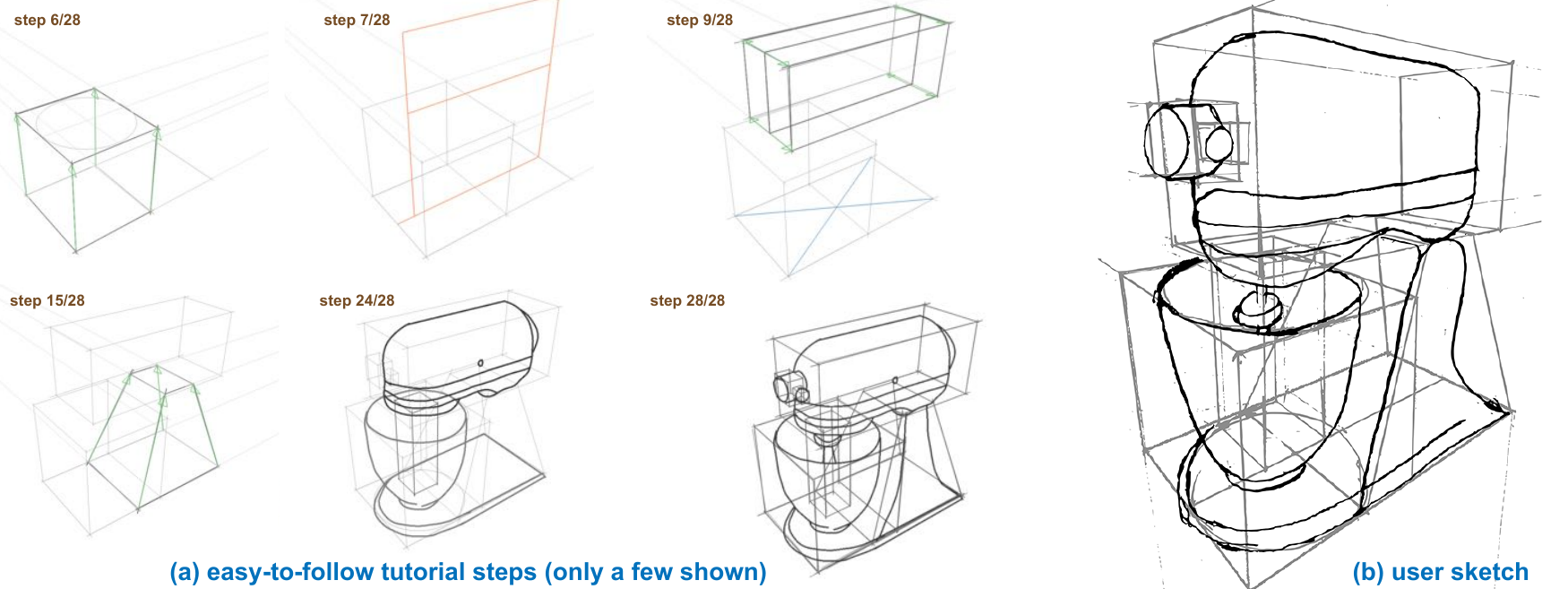}
   \caption{(a) We present How2Sketch, a system that automatically generates easy-to-follow tutorials for drawing 3D models. Each generated tutorial comes as a list of steps for drawing scaffolding primitives that help the user draw the object in correct perspective. To help the user draw the scaffolding, the tutorial shows how to construct guidelines that anchor objects parts relative to one another.  User feedback on the tutorials was positive as many created what they felt were more accurate drawings~(b) User sketch from alternative viewpoint.
   }
   \label{fig:teaser}
   \vspace{-0.15in}
 }

\maketitle

\begin{abstract}
Accurately drawing 3D objects is difficult for untrained individuals, as it requires an understanding of perspective and its effects on geometry and proportions. 
Step-by-step tutorials break the complex task of sketching an entire object down into easy-to-follow steps that even a novice can follow. 
However, creating such tutorials requires expert knowledge and is a time-consuming task. As a result, the availability of tutorials for a given object or viewpoint is limited. 
How2Sketch addresses this problem by automatically generating easy-to-follow tutorials for arbitrary 3D objects. 
Given a segmented 3D model and a camera viewpoint,
it computes a sequence of steps for constructing a drawing scaffold comprised of geometric primitives, which helps the user draw the final contours in correct perspective and proportion. 
To make the drawing scaffold easy to construct, the algorithm solves for an ordering among the scaffolding primitives and explicitly makes small geometric modifications to the size and location of the object parts to simplify relative positioning. 
Technically, we formulate this scaffold construction as a single selection problem that simultaneously solves for the ordering and geometric changes of the primitives. 
We demonstrate our algorithm for generating tutorials on a variety of man-made objects and evaluate how easily the tutorials can be followed with a user study. 

\end{abstract}

\section{Introduction}


The ability to draw real-world objects is a useful and important skill across many disciplines.  Product designers draw daily as they generate and refine product ideas, fine artists may spend hours in figure drawing classes learning how to replicate a shape from the real world, while hobbyists use sketches for visual expression.  Still, sketching requires skill and practice. One of the major challenges in drawing real-world objects is learning to draw \emph{what you see} rather than \emph{what you know}~\cite{edwards1999}. A simple cylinder, for example, is \emph{known} to have a circular cross-section with equal widths at the top and bottom. 
However, when we actually \emph{see} a cylinder, 
it is subject to perspective distortion: circles become ellipses while projected radii diminish with distance from the viewer. 

Art books and tutorials provide step-by-step instructions to teach sketching~\cite{tizon2007art,eissen2011sketching}. To help with proportions and relative positioning, they introduce geometric constructions, such as scaffolding primitives, and follow a coarse-to-fine approach, using prior steps as basis for subsequent ones. 

Authoring such tutorials requires significant expertise and time commitment even for trained artists. Consequently, objects and viewpoints in existing tutorials tend to be limited and are chosen by the expert, rather than the users of the tutorials. To address these issues, we present an approach for \emph{automatically generating easy-to-follow tutorials for drawing 3D models from user specified viewpoints.} Figure~\ref{fig:teaser} shows parts of a tutorial generated by our system 
 and the drawing by one of our study participants based on that tutorial from an alternative viewpoint. Our algorithm targets man-made objects where part relations and proportions tend to be crucial for accurate depiction.

\begin{figure}[b!]
\vnudge 
\centering
\includegraphics[width=\columnwidth]{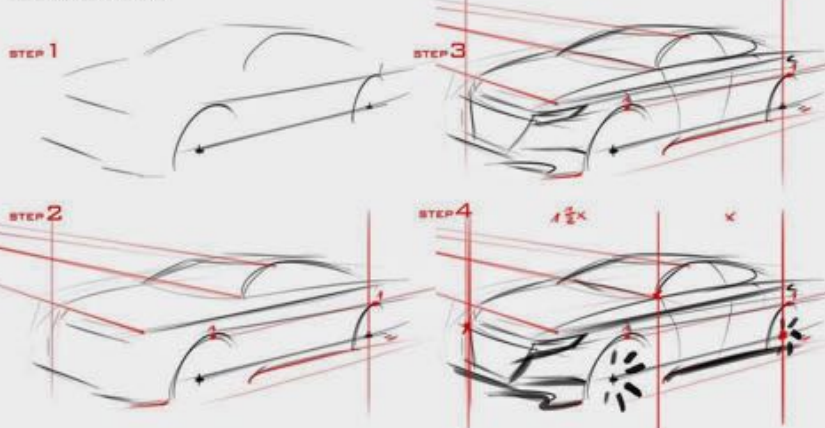}
\caption{A step-by-step sketching tutorial for drawing a car~{\small \copyright}Czajkowski. The task is made simpler by breaking it into steps and by providing guidance about part proportions and alignments. }
\label{fig:motivation}
\end{figure}

Inspired by instructional books and online tutorials, we take explicit steps to make a sketching tutorial {\em easy-to-follow}: 
%
(i)~focus on accurate inter-part proportions and {\em relations} via a drawing scaffold, followed by detailing of the object contour;
(ii)~proceed in a {\em coarse-to-fine} fashion, where object parts are abstracted as primitives (e.g., cuboids, cylinders) over several levels of detail to build up said scaffold; 
(iii)~propose a particular drawing {\em order} among the scaffolding primitives such that those sketched later can be easily anchored (i.e., drawn with guidance) off already drawn primitives; and
(iv)~provide explicit steps for the construction of {\em guidelines} to accurately anchor the scaffolding primitives.
%
%

One of our key observations is that in easy-to-follow tutorials the dimensions and arrangements of object-parts tend to have ratios that are easy to construct. For example, it is easier to construct the center line of a rectangular face compared to its one fifth line. 
Tutorial authors choose to construct with such `easy ratios' to simplify the drawing process and to focus on the procedure, rather than incidental and arbitrary measurements (see Figure~\ref{fig:motivation}). To apply this technique to existing objects, our system How2Sketch proposes small geometric changes 
while keeping overall deviations from the source model minimal. Since in each step new primitives and guidelines are anchored with respect to those drawn in previous steps, the ordering of steps significantly affects the simplicity of ratios that can be employed, and the geometric approximations that this incurs. 
This tight interdependence between ordering of primitives and their geometric changes makes the problem non-trivial. 
A further challenge is to preserve the original inter-part relationships of objects, even under geometric perturbations. For example, in Figure~\ref{fig:teaser} the coaxial relationship between the mixer bowl and mixer blade is preserved. 
%

Technically, we map the geometric adjustment and ordering of parts to a single selection problem. We first generate a set of potential candidate primitives by enumerating different anchoring possibilities. Since anchoring requires drawing guidelines, and some guidelines are easier to construct than others, the algorithm prefers anchoring possibilities that rely on easy-to-construct guidelines, such as the top edge, bottom edge, center line, etc., of existing primitives. 
Our key insight is that the problem of geometric adjustment and ordering of parts can be simultaneously solved by {\em selecting} an appropriate subset from the candidate primitives, in order to balance between geometric changes and ease of constructing necessary guidelines.
%

We test our algorithm on a range of examples and evaluate our algorithmically generated, easy-to-follow tutorials with a user study. 

\section{Related Work}

\mypara{Assisted drawing} Various applications have been proposed to assist a user in sketching. Some correct the user input based on geometric analysis of the users input strokes \cite{Igarashi:2001,Bae:2008,Schmidt:2009}, others rely on an underlying image to guide the user \cite{IBT13,xie2014,Benedetti:2014,Fernando:2014}, and yet others on crowdsourced data (e.g., many sketches) to improve the users drawing \cite{Dixon:2010,Lee:2011,Gingold:2012,Limpaecher:2013,SimoSerraSIGGRAPH2016}. These methods aim to improve the quality of the users' strokes at a local level. 
Our focus is on suggesting a meaningful drawing order and easy-to-construct guides for accurate depiction of perspective and proportions. Stroke correction or beautification is orthogonal to our main contribution and may be used to complement the contour drawing phase of our tutorials. Other assisted sketching systems take as input 2D sketches and interpret them as 3D curve networks \cite{Xu:2014:True2Form}. More advanced methods~\cite{CrossShade:SBSS12,IBB15:BendFields,Pan:2015:FlowSurfacing} use {2D} input to infer 3D geometry or surface normals for complex shading. 
We focus on the automatic generation of sketching tutorials, rather than automatic inference based on the sketched curves. 


\mypara{Tutorials} 
A good tutorial greatly facilitates understanding. Many attempts have been made to automatically generate high-quality tutorials for different applications. 
A digital drawing tutorial system was proposed by Fernquist et al.~\shortcite{Fernquist:2011} that allows an expert to create tutorials for novices. Tutorial generation systems~\cite{takagi2003learning,Cummmings2012} for specific sketching tasks 
have also been proposed, for example drawing a single scene with pre-defined objects, or `eyes.'
Grabler et al.~\shortcite{Grabler:2009} developed a tutorial system for photo manipulation tasks. 
How2Sketch produces sketching tutorials for man-made objects automatically, rather than relying on an experienced artists to create them.

%

\mypara{Drawing expertise} Tchalenko~\shortcite{Tchalenko2007} found that novices and professional artists have comparable accuracy when perform basic line drawing tasks (straight lines and simple curves). However, in a follow-up study~\cite{Tchalenko2009-seg}, he showed that when copying complex artworks, novices made significantly more errors than artists.  The main difference in drawing strategy was that experts divided complex lines into easy-to-draw short segments. Schmidt et al.~\shortcite{Schmidt2009:Experts} found that experts made qualitatively similar errors to non-artists, indicating that perspective drawing is hard, even for trained users. Particularly for off-axis viewing angles, drawing error increased significantly.  In an observational study,  Grimm~\shortcite{Grimm2011ResultsOA} found that artists commonly used a coarse-to-fine strategy starting with blocking shapes and finishing by drawing detailed items at the end. 
How2Sketch assists the user by breaking the drawing process up into basic steps that are easy to execute and by explicitly indicating vanishing line directions.


\mypara{Line drawings} Many methods for generating stylized artistic renderings of objects have been proposed (
see \cite{Kyprianidis:2013} for a survey). We leverage stylization to visually distinguish the various line types of our tutorials (perspective lines, guides, contours, etc). Other researchers investigated which features artists typically draw to convey 3D shape \cite{DeCarlo:2003:SCF,DeCarlo:2004:IRO,Burns:2005:LDF,DeCarlo:2007:HLF}. 
We display suggestive contours computed on the optimized object 
to aid the user in adding final details to their drawing. Fu et al.~\shortcite{fu2011} and Liu et al.~\shortcite{Liu:2014} infer plausible contour ordering from 2D and 3D inputs, respectively. While the derived sequences are plausible, they are not tailored for tutorials and do not provide specific guidelines to make them easy to follow.  

\begin{figure}[b!]
\vnudge
  \centering
  \includegraphics[width=\columnwidth]{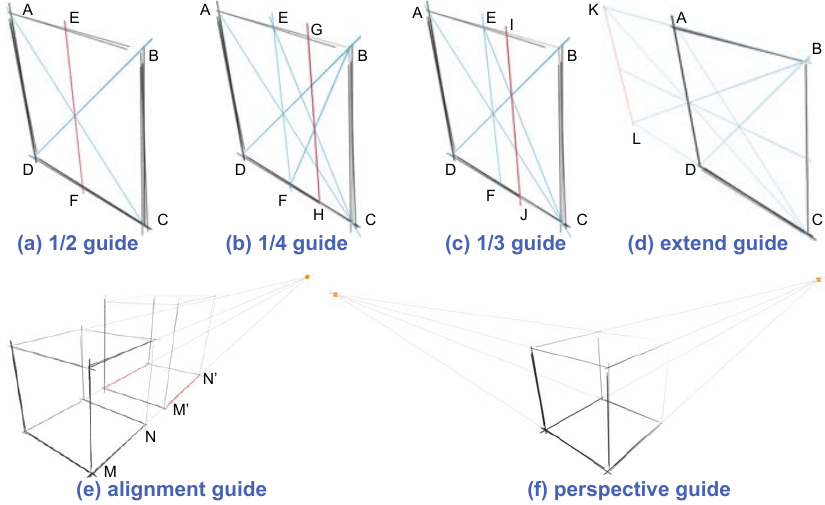} 
  \caption{Our system supports different forms of guidelines for drawing coplanar proportions~(a-d), for anchoring alignments~(e), and for previewing 2-point perspectives~(f). See Section~\ref{sec:design_rules}. }
  \label{fig:guidelines_book}
\end{figure}

\begin{figure*}[t!]
  \centering
  \includegraphics[width=\textwidth]{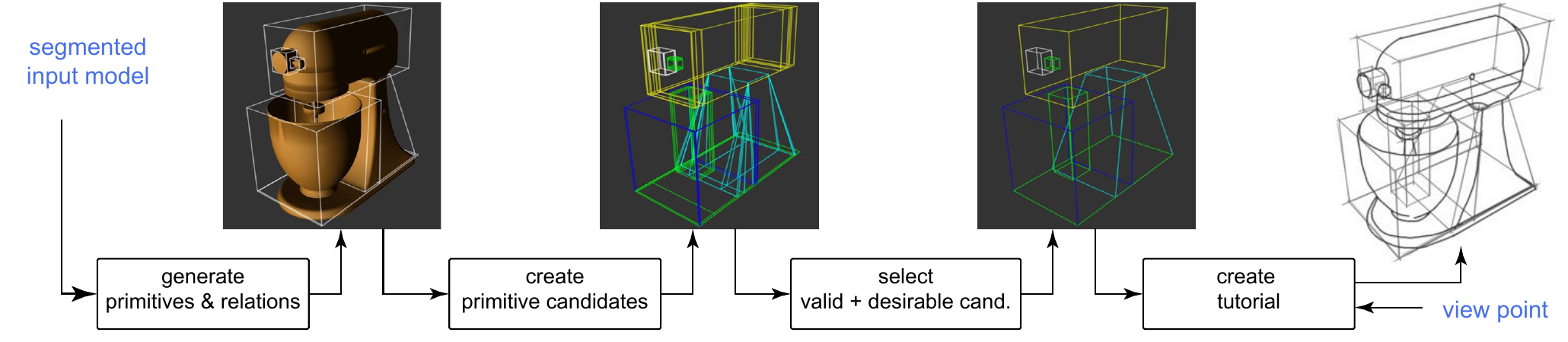} 
  \caption{System Overview. Starting from an input part-level segmented model and a user-specified viewpoint, How2Sketch generates easy-to-follow step-by-step tutorials. The system automatically makes subtle 
  geometric modifications to simplify the resultant tutorial. }
  \label{fig:overview}
\end{figure*}

\section{Learning How to Sketch}
\label{sec:design_rules}

To inform the design of How2Sketch we studied several drawing books~\cite{edwards1999,eissen2007sketching,eissen2011sketching}, visited a number of sketching websites (e.g., Sketch-a-Day), carried out an expert interview with a professional artist, and took a drawing course.

Through this process we found that effective tutorials for drawing 3D objects typically include the following: 
\begin{itemize}
\item \emph{Parts are approximated by geometric primitives:} Plane, cubes and cylinders are heavily used to approximate shapes. They are easy to construct and verify visually.
%
\item \emph{Steps are coarse-to-fine:} First, the overall object is scaffolded with approximate shapes, and then, finer contour details are added. Primitives are drawn sequentially, in a particular order.  
\item \emph{Anchor shapes to each other:} Shapes are drawn with respect to previously drawn shapes, to aid with correct placement and proportions. Instructions for positing shapes relative to each other use simple measurements (e.g., draw box \emph{half way} down the side, draw circle in the \emph{center} of the rectangle), etc. 
\item \emph{Vanishing lines for perspective:} Vanishing points are explicitly indicated to aid the user to draw correctly. 
\end{itemize}

How2Sketch supports the above tutorial features as follows:


\mypara{(a)~Scaffolding Primitives}
How2Sketch utilizes scaffolding primitives to geometrically approximate each segmented object part. The system supports planes, cuboids, cylinders, and truncated pyramids, as they allow for planar guidelines to be used, which are simple to construct, and cover a wide range of shapes. In addition to scaffolding, we guide users in drawing ellipses to better approximate some shapes. 

\mypara{(b)~Ordering}
%
Our algorithm provides the relative ordering of the scaffolding primitives. Further, How2Sketch offers detailed, sequenced instructions for constructing primitives. 

\mypara{(c)~Placement, Alignment, and Proportions}
We support a set of coplanar guidelines (see Figure~\ref{fig:guidelines_book}). Given a face $ABCD$, its diagonals help construct the \sfrac{1}{2} line $EF$ (Figure~\ref{fig:guidelines_book}a). Two levels of \sfrac{1}{2} lines produce a \sfrac{1}{4} line $GH$ (Figure~\ref{fig:guidelines_book}b);  while intersecting a diagonal $BD$ with line $CE$ produces a \sfrac{1}{3} line $IJ$ (Figure~\ref{fig:guidelines_book}c). 
Similarly, we support extrusion towards a vanishing point as in Figure~\ref{fig:guidelines_book}d where $ABCD$ is extended by reflection to form $BCLK$ such that $AB=AK$.
Finally, we also support alignment, as in Figure~\ref{fig:guidelines_book}e, $M'N'$ is aligned with $MN$. 

\mypara{(d)~Perspective}
To provide perspective information, we show the vanishing points (if within the drawing area) and also show the vanishing lines leading to them (Figure~\ref{fig:guidelines_book}f). How2Sketch supports sketching in 2-point and 3-point perspective.

\section{Generating Sketch Sequences}
\label{sec:algo}

\begin{table}[b!]
\centering
\caption{Notation table.}
\label{tab:notation} \small
\begin{tabular}{r|l}
symbol & denotes \\ \hline\hline
 $S$ & input part-segmented model \\
 $P_i$  &  primitive corresponding to the $i$-th part of $S$ \\
 $R_{i,j}$ & relation between primitive pairs $(P_i,P_j)$    \\
 $ C_{j \rightarrow i}^k$ & candidate for the $i$-th part primitive with (anchoring) parent from the \\  
    & $j$-th part primitive, where $k$ denotes the $k$-th such instance \\
    $\setcand{i}$ &  set of all the candidate primitives generated for part primitive $P_i$ \\
 $\chi(X)$ &  indicator variable corresponding to the selection of $X$ \\
 $\Lambda$ & assignment of indicator variables denoting a set of selected candidates
\end{tabular}
\end{table}

Given a 3D object ($S$) segmented into parts and a desired viewpoint, our goal is to establish an easy-to-follow sequence for drawing the object, starting with the scaffolding and progressing to the contour details. We make it easier to draw the scaffold by actively making small part-level geometric changes to facilitate relative anchoring using a set of guidelines.

As described in Section~\ref{sec:design_rules}, we have adopted simple procedures to accurately draw guidelines at easy-to-construct ratios (\sfrac{1}{2}, \sfrac{1}{3}, \sfrac{1}{4}, 1$\times$, 2$\times,$ etc). Object part placements and sizes in the original models, however, rarely conform to such ratios. Hence, we propose to modify object parts, so that they end up with part relationships that are easy to draw. We motivate this choice twofold: (i)~Scaffolding primitives in tutorials like those generated by How2Sketch are already approximations of real geometry and thus contain a measure of error. Some of this error can actually be compensated by adjusting the fit of contours within the scaffold. (ii)~Accurate estimation of lengths and ratios is difficult, even for experts, so errors are almost unavoidable. By enforcing that parts relate via simple ratios for which reasonable geometric constructs can guide the user, the overall drawing error is unlikely to increase significantly beyond the unguided case.

Our algorithm proceeds in three main stages (see Figure~\ref{fig:overview}):  
(i)~generating part-level primitives and encoding inter-primitive relations; 
(ii)~creating primitive candidates based on various inter-primitive anchorings strategies; and 
(iii)~selecting a valid and desirable set of primitives among the candidate selections. The result implicitly encodes 
how to geometrically modify each part (both their dimension and placement), and in which order to draw them. Intuitively, our algorithm produces an easy-to-follow primitive drawing sequence at the cost of deviating from the original geometry in a controlled fashion. 
We now elaborate each step. Please refer to Table~\ref{tab:notation} for symbols used in the following.

\begin{figure}[t!]
  \centering
  \includegraphics[width=\columnwidth]{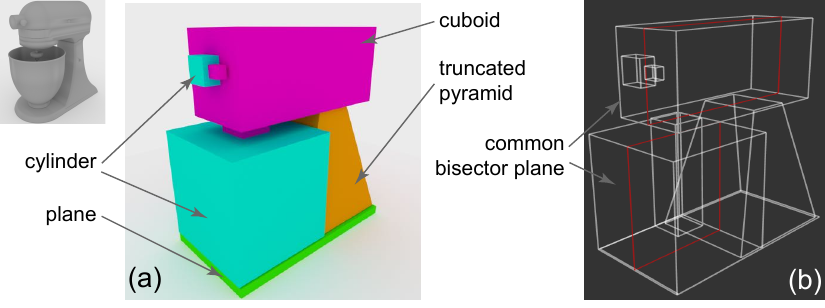}  
  \caption{Given a part-segmented input model $S$ (top-left inset), we abstract the parts as different primitives~(a) and identify inter-part relations. For example, here the mixer bowl and mixer head primitives share a common bisector plane. }
  \label{fig:primitive_types}
\vnudge
\end{figure}

\subsection{Generating Primitives and Inter-part Relations}

We abstract model parts by primitive shapes. In our implementation we support planes, cuboids, cylinders, and truncated pyramids (see Figure ~\ref{fig:primitive_types}a). (Note that in our visualization we show axis-aligned bounding box for cylinders as the box faces are used for providing guidance for drawing ellipses.) For each part of the input model $S$, we use least-squares to fit (axis-aligned) different primitive types and take the one with the least residue. In case of ties, we prefer the simpler primitive. 
We denote the primitive for the $i$-th part as $P_i$ (type of primitive is not explicitly indicated in this notation).

Man-made objects, which are our target objects, often have dominant inter-part relations. We found it highly desirable to preserve such relations in the generated tutorials. Hence, we first detect such inter-part relations and later preserve them in the generated tutorials. 
We simply test (see \cite{mzlsgm_abstraction_siga_09}) each pair of primitives $P_i$ and $P_j$ for any relations. In our implementation, we consider coplanar, coaxial, and common bisector plane relations. In case of multiple relations between a pair of primitives, we prefer common bisector plane over coaxial over coplanar. 
We represent a relation using a binary variable $R_{i,j}$ where $i$ and $j$ respectively denote the primitives $P_i$ and $P_j$ (type of relation is not explicitly indicated in this notation). 
If a relation is present, we mark $R_{i,j} = 1$, and $R_{i,j} = 0$ otherwise. 
Figure~\ref{fig:primitive_types} shows some examples.


\begin{figure}[t!] 
  \centering
  \includegraphics[width=\columnwidth]{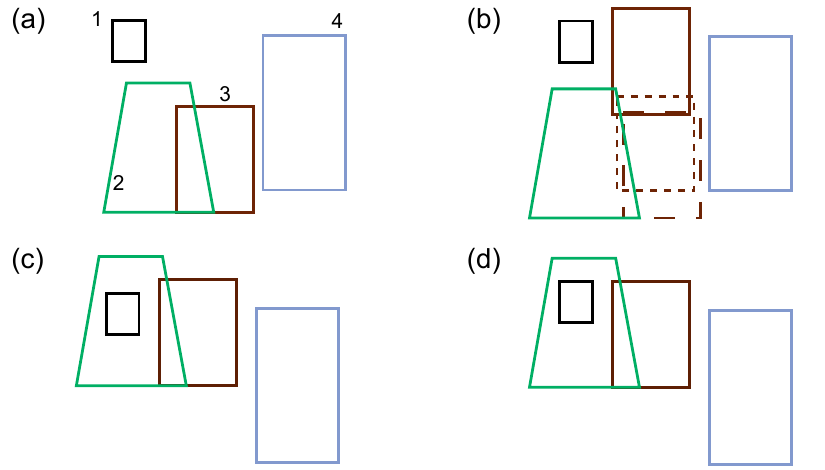} 
  \caption{Starting from initial primitives $P_1, P_2, P_3, P_4$, for each pair of primitives we generate several candidate primitives. For example in (b), 
  we show the primitives generated for $P_3$ using $P_4$ as parent, indicated as different instances $P_{4\rightarrow 3}^k$ shown in different dotted/solid brown.  
  Relations are restored leading to further new primitives, for example in (c) the green primitive was lifted to restore coplanarity with the brown primitive. 
  Finally, we also have second level primitives. As shown in (d), the new brown primitive leads to a new primitive for $P_1$. 
  This is an illustrative figure in 2D with only some of candidate primitives shown for simplicity. 
  }
  \label{fig:candidate_generation}
  \vnudge
\end{figure}

\subsection{Creating Candidate Primitives}
We now describe the candidate primitive generation step that creates additional primitives based on possible anchoring strategies. We use $\setcand{i}$ to denote the set of all the candidate primitives generated corresponding to primitive part $P_i$. Since the the original primitive is always a candidate, we start by $\setcand{i} := \{P_i\}$. We generate candidate primitives in three stages: 

(i)~ For each pair of primitives $P_i$ and $P_j$, we generate candidates of the form $ C_{j \rightarrow i}^k$, where $j \rightarrow i$ indicates that a candidate is generated for primitive part $P_i$ and is anchored off $P_j$ with $k$ denoting different anchoring possibilities. For example, parts can be anchored based on different guidelines  described in Section~\ref{sec:design_rules} for different face- or plane-based anchors. We append these candidates to the respective candidate sets as: 
$\setcand{i} \leftarrow \setcand{i}  \cup \{C_{j \rightarrow i}^1, C_{j \rightarrow i}^2, \dots \}$ (see Figure~\ref{fig:candidate_generation}b). 
(ii)~For each pair of primitives $(P_i, P_j)$ sharing a relation of the form $R_{i,j}$, we add additional primitives to their candidate sets to restore the relations. Specifically, corresponding to 
a candidate of the form $C_{j \rightarrow i}^k$ (created in stage (i)), we create a new candidate of the form 
 $C_{i \rightarrow j}^{k'}$ such that $ C_{j \rightarrow i}^k \leftrightarrow C_{i \rightarrow j}^{k'} $ are similarly related as in $P_i \leftrightarrow P_j$. 
We append all such relation-based additional candidate primitives to the respective candidate sets, i.e., $\setcand{i} \leftarrow \setcand{i}  \cup C_{i \rightarrow j}^{k'}$ (see Figure~\ref{fig:candidate_generation}c).

Note that in the above a candidate is allowed to be anchored from one or multiple parents, as each axis can be independently anchored. Additionally a candidate can be partially unguided (e.g., the width and length of cuboid is guided but the height is not) or completely unguided (e.g., it is simply the input primitive) (see Figure~\ref{fig:candidate_generation}).  
We defer further details to the implementation section.

(iii)~We allow second-level anchors, i.e., candidate primitives as generated above are allowed to act as anchors for other primitives. To this end, we simply iterate one more time stage (i) and (ii) (see Figure~\ref{fig:candidate_generation}d). Note that before starting this step, we
remove the candidate primitives with large changes in geometry or relative placements (more details in the implementation section). 

At the end of this stage, we have a set of candidates for each part of the input model resulting in the super set of candidate primitives of the form $\{ \setcand{i} \}$ (see Figure ~\ref{fig:primitive_selection}).

\subsection{Selecting Candidate Primitives}

Having generated multiple candidates, our remaining task is to select a set valid and desirable candidates, as explained next.

\mypara{Valid candidate sets} We first characterize the notion of valid selections. 
We use indicator variables $\chi(X)$ to denote if a candidate primitive $X$ is selected (i.e., $\chi(X)=1$) or not (i.e., $\chi(X)=0$). We have $ \chi(C_{j \rightarrow i}^k) \in \{0,1\}$ for each $C_{j \rightarrow i}^k \in \setcand{i}$. 
Let $\Lambda$ denote a particular assignment for the indicator variables for {\em all} the candidate primitives.

Among the various possible selections, not all the subsets of candidates of the form $\Lambda$ constitute {\em valid} selections. A valid selection of candidates should satisfy three conditions: \\
(1)~for each part of $S$, only {\em one} candidate primitive should be selected; \\
(2)~if a selected candidate primitive is anchored off one or more parent (candidate) primitives, then its parent primitive(s) {\em must} also be selected; \\
(3)~if any two primitives $P_i$ and $P_j$ share a relation, then their corresponding selected candidate primitives should also respect the same relation.  (Note that by (1) implies that each primitive should have a unique selected candidate associated with it.)

We now express the above conditions in terms of the indicator variables in $\Lambda$.  \\
(a)~We encode (1) as  
\begin{equation}
\sum_{j,k} \chi(C_{j \rightarrow i}^k) = 1 \quad\quad \forall i.
\label{eqn:selectOne}
\end{equation}
(b)~We encode (2) as a quadratic constraint involving the binary selection variables as 
\begin{equation}
\chi(C_{j \rightarrow i}^k) \chi(C_j) - \chi(C_{j \rightarrow i}^k) \ge 0
\label{eqn:validSelect}
\end{equation}
for each dependent pair $C_{j \rightarrow i}^k \in \setcand{i}$ and its parent $C_j$. 
Note that this condition {\em disallows} $\chi(C_{j \rightarrow i}^k)=1 \textsc{ AND } \chi(C_j)=0$, but allows any of the other three assignments involving $\chi(C_{j \rightarrow i}^k)$ and $\chi(C_j)$. \\
(c)~We now encode (3). Let two primitives $P_i$ and $P_j$ share a relation, i.e., $R_{i,j}=1$. 
Let $\setcand{i} = \{C_{*i}^1, C_{*i}^2, \dots \}$ be all the generated candidates for primitive $P_i$ and similarly  $\setcand{j} = \{C_{*j}^1, C_{*j}^2, \dots \}$ for primitive $P_j$. Then for {\em each}  pair of the form
 $C_{*i}^{k} \in \setcand{i}$ and $C_{*j}^{k'} \in \setcand{j}$ that does {\em not} share the same relation as $R_{i,j}$, we require
\begin{equation}
\chi(C_{*i}^{k}) \chi(C_{*j}^{k'}) = 0.  
\label{eqn:invalidRelation}
\end{equation}
This condition disallows $\chi(C_{*i}^{k})=1 \textsc{ AND } \chi(C_{*j}^{k'})=1$, i.e., they cannot be jointly selected as these candidate primitives do not share the same relations as of their respective primitive parts. 

\begin{figure}[t!]
  \centering
  \includegraphics[width=\columnwidth]{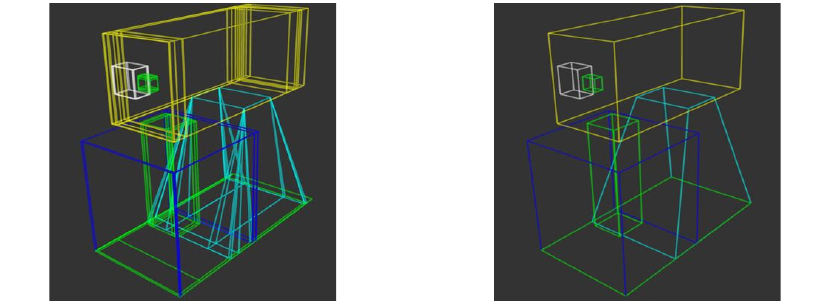} 
  \caption{From a set of candidate primitives~(left), our algorithm selects a subset of primitives that is {\em valid} and {\em desirable} as shown on the right. 
  The selection implicitly encodes in which order to draw the primitives and also how to change each primitive (size and/or placement) such that the resulting tutorial is easy to construct. 
  Please refer to the text for details. }
  \label{fig:primitive_selection}
\vnudge
\end{figure}

Thus, a selection $\Lambda$ is valid if Equations~\ref{eqn:selectOne}-\ref{eqn:invalidRelation} are all satisfied. 
Among all such valid selection sets, we next determine which one is most desirable. Figure ~\ref{fig:primitive_selection} shows a set of candidate primitives and a valid selection.

\mypara{Sequencing sketching as a selection problem} 
We balance the error due to making changes to the geometry with difficulty of drawing arising from anchoring. In other words, an unanchored part is more difficult to draw compared to an anchored part -- this is a key hypothesis of this work. We indicate this difficulty of drawing cost as $E_e(C_{j \rightarrow i}^k)$ with a lower cost denoting easier to draw (see later for details). The total cost is expressed as: 
\begin{equation}
E_{\text{difficulty}}(\Lambda) := \sum_{i,j,k} \chi(C_{j \rightarrow i}^k) E_e(C_{j \rightarrow i}^k). 
\end{equation}
Selecting any primitive, however, incurs an associated error that we indicate as $E_d(C_{j \rightarrow i}^k)$ due to deviation from original geometry (see later describe how we measure $E_d$). So, the total data 
cost of selecting a set of primitives is: 
\begin{equation}
E_{\text{adjust}}(\Lambda) := \sum_{i,j,k} \chi(C_{j \rightarrow i}^k) E_d(C_{j \rightarrow i}^k)
\end{equation}
with a higher cost indicating larger geometric deviations from the original parts.

Thus, we arrive at the final formulation for {\em desirable} selection as,
\begin{equation}
\min_{\Lambda} (E_{\text{adjust}}(\Lambda) + E_{\text{difficulty}}(\Lambda)) 
\label{eqn:formulationBasic}
\end{equation}
subject to Equations~(\ref{eqn:selectOne})-(\ref{eqn:invalidRelation}) to ensure a valid selection. Thus, we have formulated our problem as a quadratically constrained linear program.

\mypara{Error functions}
The above formulation requires metrics for $E_e$ and $E_d$. While various metrics may be substituted, we used the following in our implementation. 

For the difficulty of drawing term $E_e(C^k_{j \rightarrow i})$, we associate a higher cost for anchors that are harder to replicate (e.g., requiring more construction lines). Specifically, we set the cost to the number of guidelines divided by the  area of the parent plane where construction lines are to be drawn. This encourages fewer guides but also using planes/faces with larger areas for drawing sketch guides. (The effect of viewpoint is only considered at runtime as discussed in Section~\ref{sec:interface}).

For the data error $E_d(C^k_{j \rightarrow i})$, we sum the changes in length along each axis, normalized by the original axis length, with the translation of the midpoint of each axis, again normalized by the input axis length. For an unguided axis we set the data error to the maximum of $2$ to discourage unguided candidates.

\mypara{Final drawing order}
The solution to the above optimization directly gives us both the {\em ordering} and the {\em modifications} of the parts. 
Note that the above solution may only return a partial ordering among the candidates primitives. This implies that the relative drawing order among of certain primitives are not specified. We break such ties only at runtime once the user selects a view as described in Section~\ref{sec:interface}. 
%

{\em A note about greedy alternative:} An alternative algorithm is to greedily consider primitive pairs and snap them relative to nearest guidelines. 
%
Such an adhoc strategy fails to take care of the second level candidates, and produces an inferior solution. More importantly, this approach does not provide any meaningful ordering based on ease of drawing, which is a key focus of our problem.

\subsection{Implementation details}
We now clarify some additional implementation details. Most of these are choices we made in our implementation, and can be replaced by multiple comparable alternatives. 

Primitives that are not well approximated by one of the currently supported ones can be represented as a custom primitive (e.g., line) but such primitives cannot be part of our optimization step. Instead after our optimisation step their relative positioning is updated and when sketched they are unguided. Alternatively, the user can approximate them by their bounding box primitive. 

The candidate primitive generation works in two steps: first, we use the coplanar relations to generate candidate planes $ c_{i \rightarrow j}^k$, and then depending on the primitive type we combine the planes to create a complete primitive $C_{i \rightarrow j}^k$ (here,  lowercase c for candidate plane rather than complete primitive that uses uppercase C). This choice unifies candidate primitive generation across primitive types (recall cylinders are processed based on their axis-aligned bounding box).

For each pairwise coplanar relation $R_{i,j}$ we have two participating planes in $P_i$ and $P_j$: at this stage the relation is undirected and we produce candidate planes using both combinations $ c_{j \rightarrow i}^k$ and $ c_{i \rightarrow j}^k$. To generate a candidate plane, each axis is considered independently then all combinations are combined to create planes $ c_{i \rightarrow j}^k$. An axis can be anchored by the parent plane using the end points of the same axis. This means there are several anchoring possibilities.  For example, anchoring the vertical axis of $P_i$ on $P_j$ might involve anchoring the top edge of $P_i$  to the \sfrac{1}{3} line of  $P_j$ and the bottom edge $P_i$ to the bottom edge of $P_j$. An alternative might be to anchor  the top edge of $P_i$  to the \sfrac{1}{3} line of  $P_j$  and the bottom edge $P_i$ to \sfrac{1}{4} line of $P_j$. We initially generate all candidates but to reduce the number of candidates to select from we discard those where an axis length or translation change of more than $10\%$ of the input length. 

With all the candidate planes generated using all the pairwise relations we generate complete primitives by combining the different planes based on the primitive type. To generate a complete cuboid primitive, for example, we find the missing height axis from one of the other planes to complete the primitive. For truncated pyramids we combine top and bottom planes with a height axis to make a truncated pyramid. To generate second level candidates, we repeat this process but use the first level candidates as the parent primitives. 

We use the Gurobi linear Solver~\cite{gurobi} to solve the quadratically constrained LP as described above. Typically the solver takes 1-2 minutes in the presented examples.

\begin{figure}[b!]
\vnudge
   \includegraphics[width=\columnwidth]{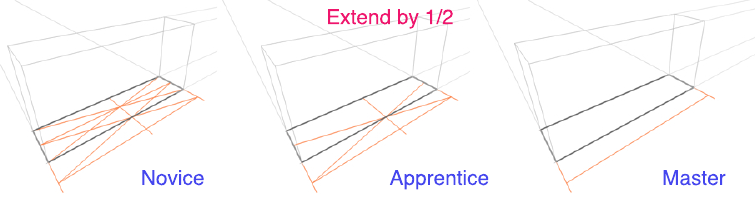}
   \caption{User ability. Users specify drawing level (novice/apprentice/master) which determine the number of intermediary guides presented for each step. For the `extend by 1/2' step, novices (left) are shown 9 guidelines, apprentices~(center) 6 guidelines, and masters~(right) 3 guidelines. }
   \label{fig:user_ability}
\end{figure} 

\section{Presenting Sketch Sequences}
\label{sec:interface}

The sequence generated in Section \ref{sec:algo} provides primitive ordering, sketching guidelines, and adjusted part geometry for drawing the scaffolding of the object. 
How2Sketch tutorials can be adapted further based on the user chosen viewpoint and user indicated drawing level (novice/apprentice/master), which can be controlled interactively. 
Our custom viewer indicates when guidelines can be erased and provides hints for drawing in perspective and object contours.

\mypara{Viewpoint} We use the specified viewpoint to customize the tutorial as follows: 
(i)~Although primitive ordering is determined based on anchoring strategies, multiple primitives can anchor from the same parent, resulting in a tie. We break such ties in ordering by first choosing the primitive that is closest to the user from the indicated viewing position. 
(ii)~The selected viewpoint can make some guidelines cumbersome to draw because of limited space on the projected area of a primitive face. We identify such instances by thresholding based on $\text{A}_p/k$, where $\text{A}_p$ indicates the projected area and $k$ the number of guides necessary to draw the primitive. If a primitive falls below this threshold, we ask the user to simply `eyeball' the primitive without drawing intermediate guides. 
(iii)~Finally, a segment that is occluded and its primitive does not help anchor any other visible primitive is deemed unnecessary and hence is left out from the tutorial.

\begin{figure}[t!]
   \includegraphics[width=\columnwidth]{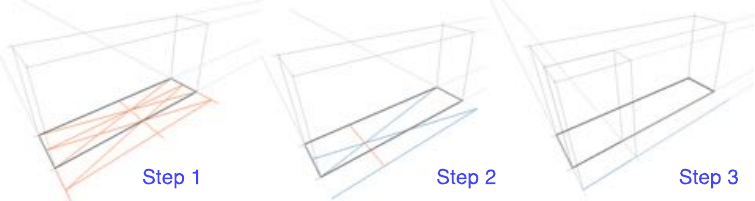}
   \caption{Guide lifetime. Guides first appear in orange (left). In subsequent steps guides that are no longer required are removed, while those that are to be reused are marked in blue (middle, right). }
   \label{fig:guide_lifetime}
   \vnudge 
\end{figure} 

\mypara{User ability} We adapt our tutorials to different sketching abilities by classifying the various guidelines as suitable for novice, apprentice, or master users. For example, dividing a face of a primitive into halves requires three guidelines. A novice is shown all the three, an apprentice only the \sfrac{1}{2} line itself, and a master is not provided with any intermediate guidance. Note that in all cases, the user is instructed to divide the highlighted face into half by a text label in the viewer (see Figure~\ref{fig:user_ability}).

\mypara{Guide lifetime} In order to reduce the amount of guidelines on a sketch at any point in time, we determine each guide's lifetime to inform the users when a guide can be safely erased. To this end, we first go over the list of generated guidelines to identify the equivalent ones, and store their {\em lifetime}, i.e., when they first appear and when they are last used. 
During the tutorial, a guideline is drawn in orange when it first appears. If the guideline is used in any later step, it is changes to blue. After the last step a guide is used, it is no longer shown. As a result, users do not have to unnecessarily erase/redraw guides, and thus reduce clutter as they sketch (see Figure~\ref{fig:guide_lifetime}).  

\mypara{Vanishing points and ellipses} Vanishing lines and vanishing points are indicated with respect to the paper boundary (shown as green corners) to help users better position the lines. We additionally guide users in sketching ellipses on a primitive face by using guides to the vanishing points. These guides intersect with the edges of the face at the perspective mid-points, which are the points where the ellipse should touch the face of the primitive. 

\mypara{Contour ordering} Once the user has sketched the scaffolding and ellipses, we guide them to sketch the contours. We use suggestive contours~\cite{DeCarlo:2003:SCF} computed on the modified underlying model segments (the contours are computed at a segment level). The contour segments are progressively displayed per segment, following the order determined by the primitives. Already drawn parts are used to determine occlusion for the new primitives, thus reducing clutter (see Figure~\ref{fig:teaser}). 


\mypara{Interface} How2Sketch tutorials can be presented in many different forms. They can be navigated manually using an our interface; they can be printed (see supplementary material), or sequenced into a video.  


%

\begin{figure*}[t!]
   \includegraphics[width=.95\linewidth]{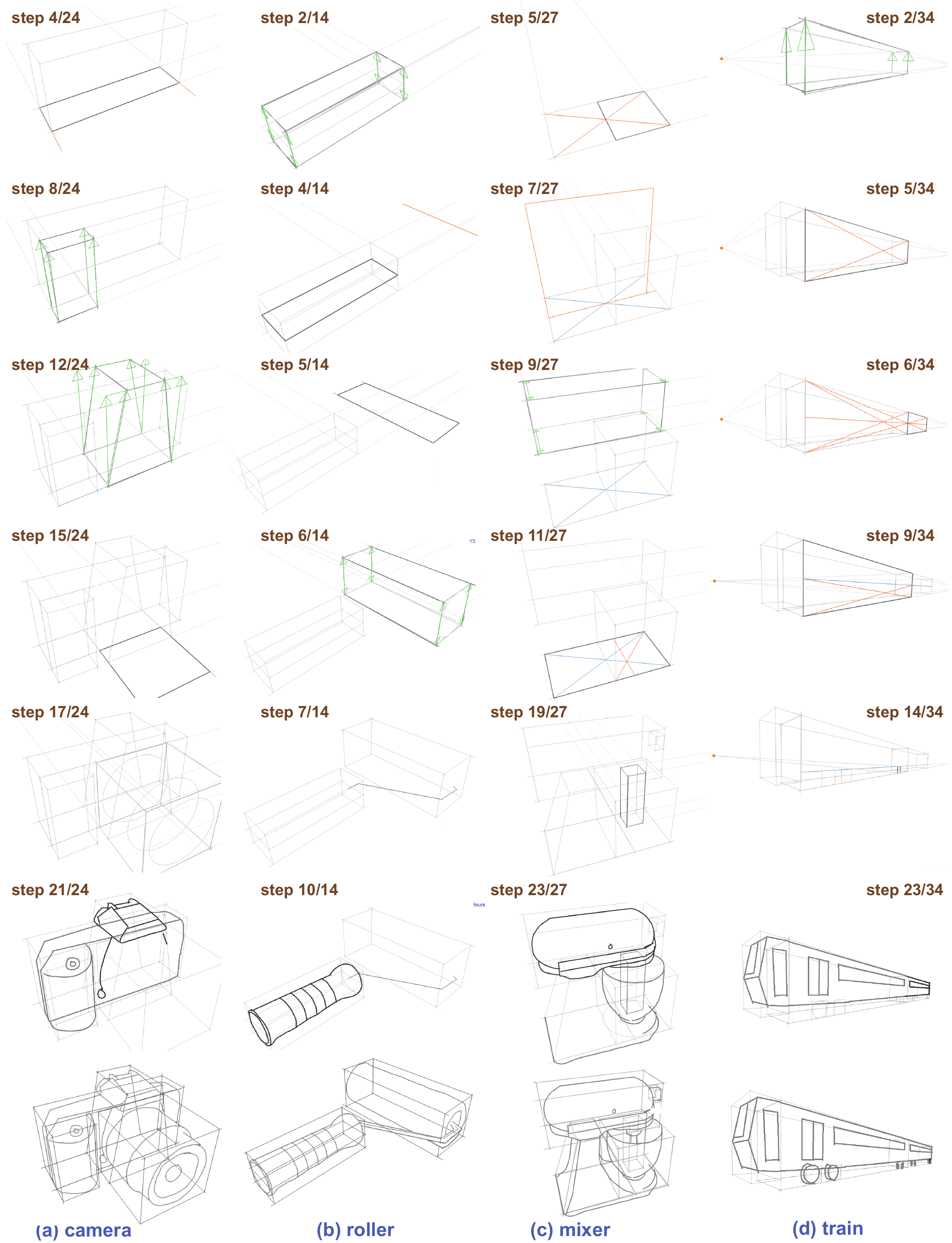}
   \caption{Example step-by-step tutorials generated by our system: (a) and (b) were generated in the master-user setting, while (c) and (d) were generated in the novice-user setting. 
   Please refer to the supplementary materials for complete examples. }
   \label{fig:results_plate}
\end{figure*}

\section{Results and Discussion}

\begin{figure}[t!]
   \includegraphics[width=\columnwidth]{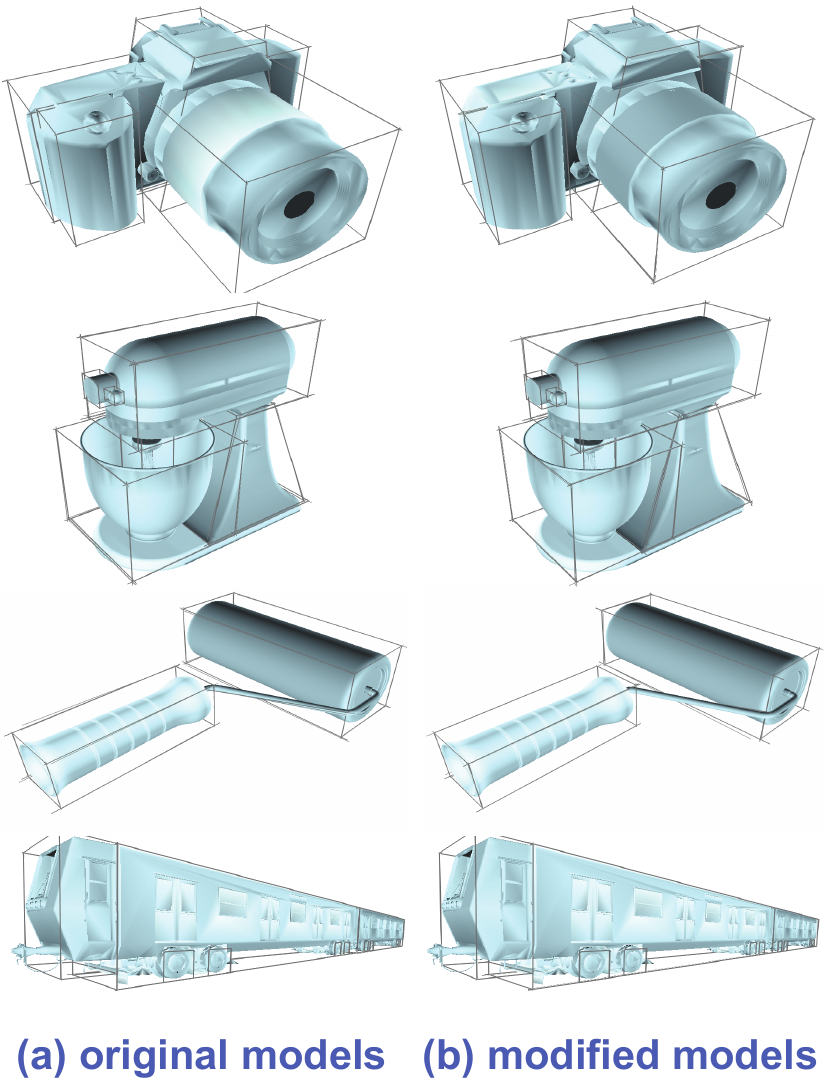}
   \caption{(Left)~Original models.  (Right)~Subtle changes proposed by our algorithm in order to make the objects easier to draw. }
   \label{fig:original_deformed}
   \vnudge
\end{figure}

We used How2Sketch to generate sketching tutorials for four man-made objects - a Digital SLR Camera, Kitchen Mixer, Train and Paint Roller. For these models, numerous tutorials depending on viewpoint and user ability can be generated. Parts of the tutorials are shown in Figure~\ref{fig:results_plate} (see supplementary material for full sequences). Each tutorial takes between 15 and 45 minutes to complete due to their varying complexity. The small changes made to the input geometry by the method are illustrated in Figure~\ref{fig:original_deformed}. As desired, the alterations to geometry are subtle but now enable simple anchoring strategies based on the altered segment bounding boxes (also shown).

As demonstrated in Figure~\ref{fig:results_plate}, our tutorials follow a coarse-to-fine strategy, starting with a single primitive that can be used to anchor subsequent primitives. Figure~\ref{fig:results_plate}a shows excerpts from a tutorial sequence with master user ability. Here, the grip is anchored on the edges of the camera body and a \sfrac{1}{4} guide. Additionally, the grip and flash are both extended by one half the depth of the main body. The lens, an example of a second level anchoring, uses the flash  for anchoring by extruding 1$\times$. Guides for ellipses are provided before contours are drawn. 

In the paint roller tutorial in Figure~\ref{fig:results_plate}b the handle anchors the roller using the common bisector plane. The top edge of the roller is 1$\times$ the length of the handle. The bottom edge is \sfrac{1}{2}$\times$ the length of the handle but due to the limited projected area and number of guides otherwise required, the step is unguided (as per Section~\ref{sec:interface}). 


Figures~\ref{fig:teaser} and \ref{fig:results_plate}c both show novice ability tutorials for the food mixer but from different viewpoints. The plane primitive for the base of the mixer anchors the bowl using a planar relation and \sfrac{1}{2} guide. The common bisector plane between the base and the main body of the mixer is used for anchoring the length of the main body. The bisector plane is first drawn before being extended in both directions to create the cuboid primitive. The Mixer's stand is an example of a primitive with two parents, being anchored off both the main body and base. 

The train example, Figure~\ref{fig:results_plate}d, anchors the second carriage as 1$\times$ the length of the first carriage and the top edge of the wheels using the \sfrac{1}{4} guide on the vertical axis of the first carriage. The driver's compartment is unguided.

\mypara{Limitations} How2Sketch only makes small changes to the input geometry. However, small gaps between object parts can have important semantic meaning. An example of this can be seen in Figure~\ref{fig:original_deformed} where the main body of the mixer and the stand separate slightly in the adjusted version. We know these two segments would be joined by a hinge making such an adjustment unrealistic. Symmetry or regular structure can similarly be lost from the small geometry changes. An example of this is the roller in Figure \ref{fig:original_deformed}, which ceases to be a perfect cylinder. Note that most of these violations are difficult to spot unaided and tend to get masked by drawing inaccuracies. Finally we find relations from the input segments but do not allow adjustments in geometry to create a relation that was not already present. In the future, we might enable such changes to allow for an even wider range of candidates.

\section{Evaluation}

To evaluate the effectiveness of the How2Sketch tutorials, we compare to a simple step-by-step tutorial that shows scaffolding primitives for each part of the object but does \emph{not} simplify the sizes or locations of the primitives to make them easier to draw. In this Basic tutorial type, the scaffolding primitives are shown in order from largest to smallest with a base primitive anchored to the ground plane. No guidelines are shown. Please see the supplemental materials for the complete tutorials used in the study. 

\mypara{Preliminary Study} We conducted a preliminary study with 8 participants comparing an earlier version of How2Sketch with the basic tutorial type. User responses to questions about satisfaction with their drawing, perceived accuracy of their drawing, and experience with the tutorial were significantly higher for the How2Sketch tutorial compared to the Basic tutorial. Additionally an ANOVA across tutorial type and object drawn revealed a significant effect of tutorial type on satisfaction, accuracy, and experience ratings (p $<$ 0.022 in all cases). Despite this positive feedback we did not observe an improvement in drawing quality across tutorial type. Based on the preliminary study observations and user feedback, the current version of How2Sketch introduces a wider range of primitives, indicates \emph{guide lifetime}, adapts tutorials based on \emph{user ability}, and uses \emph{relations} for candidate generation.

\mypara{Participants}
We recruited 10 participants (ages 18-55+, 6 men) with varied expertise in drawing. Two participants reported never drawing. Four reported drawing once in a while. Six reported drawing at least once a month. Three had taken college-level art classes or private/non-accredited art classes. When asked (free-form) what they found most challenging about drawing, 4 mentioned perspective, proportions, scale, and relative positions. When asked to rate their drawing skills on a scale of 1 (poor) to 5 (great), only 4 people rated their drawing skills above 2. 

\mypara{Methodology}
In advance, each participant filled out an introductory online questionnaire about their experience with drawing real objects. Upon arrival, each participant was told that they will be asked to draw two objects, a camera and a mixer, using two different tutorials. Participants always followed a How2Sketch tutorial first to disadvantage How2Sketch to any learning effect. The two objects (camera and mixer) counter-balanced with half of the participants using the Basic tutorial type for the camera and half using the How2Sketch tutorial for the mixer. The study had 4 conditions (2 objects x 2 tutorial types). The How2Sketch tutorial was set to the \emph{novice} ability for all participants. 

Before the How2Sketch tutorial, participants were given a written handout (see supplemental material) that described how to draw construction lines for \sfrac{1}{2}, \sfrac{1}{4}, and \sfrac{1}{3} guidelines and extending planes (see Figure~\ref{fig:guidelines_book}). This written tutorial was designed to give them context for what they would encounter in the How2Sketch condition. 

Both the Basic and How2Sketch tutorials were followed using a Macbook Pro 13" laptop; participants used the trackpad to advance forward and backward through the tutorial. All drawings were done on paper. Each participant was given two pencils (HB, 0.3mm and 0.7mm). They were allowed to use a provided straight-edge and eraser. For creating each drawing, the participants were given a sheet of paper that included the vanishing points and the ground plane of the first primitive. This initial calibration allowed us to more easily compare drawings across users. All users drew the scaffolding primitives first on the calibrated paper. For drawing the final contours of the object, the moderator attached a transparent sheet to the paper with the scaffolding. This allowed us to more easily compare both the contour drawings and the scaffolding primitives across users.

Participant filled out a questionnaire after drawing each object, indicating their level of satisfaction with their drawing (1 - not at all, 5 - very much), perceived accuracy of their drawing (1 - not at all accurate, 5 - very accurate), enjoyment with the tutorial experience (1 - not at all, 5 - very much), and how ease the tutorial steps were to follow (1 - not at all easy, 5 - very easy). They also gave free-form responses about what they liked about each tutorial type and how it could be improved. At the end of the study, subjects were asked which tutorial type they preferred (Basic or How2Sketch). In the study itself we referred to the Basic tutorial type as the tutorial \emph{without guides} and the How2Sketch tutorial type as the \emph{tutorial with guides}. 

All participants were given a \$25 gift card for their time.

\begin{figure}[t!]
\begin{overpic}[width=\columnwidth]{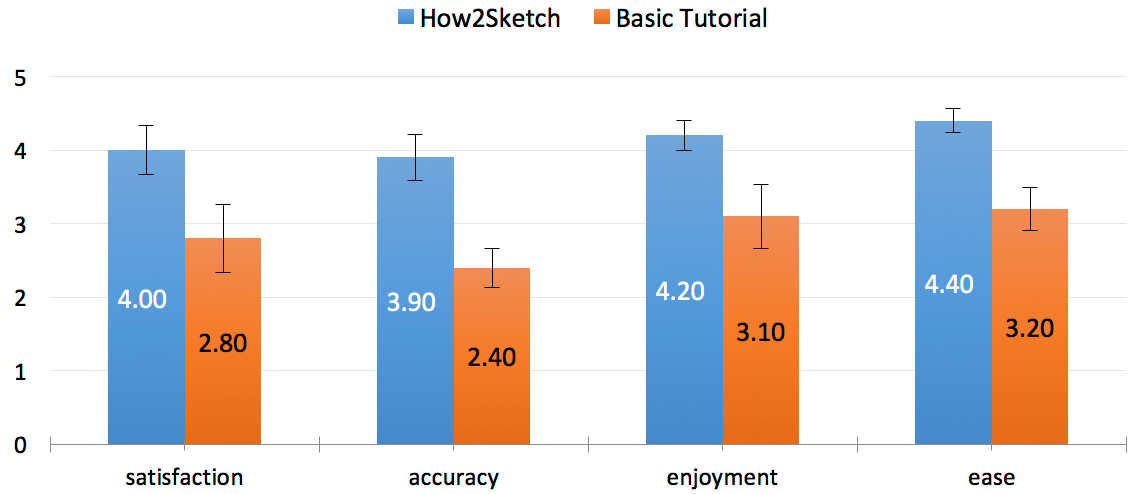}
\end{overpic}

\caption{Average user ratings for satisfaction, perceived accuracy, enjoyment, and ease of drawing were all higher for the How2Sketch tutorials than for the basic tutorials. Showing standard error of mean~(SEM) bars for $N=10$.}
\label{fig:survey}
\vnudge
\end{figure}

%

\begin{figure}[h!]
\begin{overpic}[width=\columnwidth]{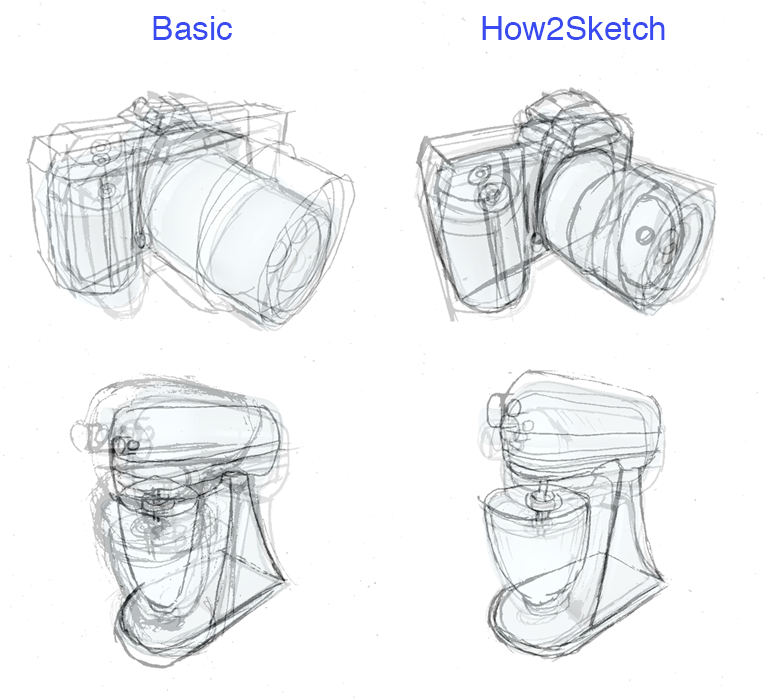}
\end{overpic}
\caption{User Study Sketches: User sketches all overlaid on the object they drew sketches from following basic tutorial (left) show much greater variation in proportions and alignment than sketches from following How2Sketch tutorials (right).}
\label{fig:sketch_overlay}
\vnudge
\end{figure}

\begin{figure}[b!]
\vnudge 
\begin{overpic}[width=\columnwidth]{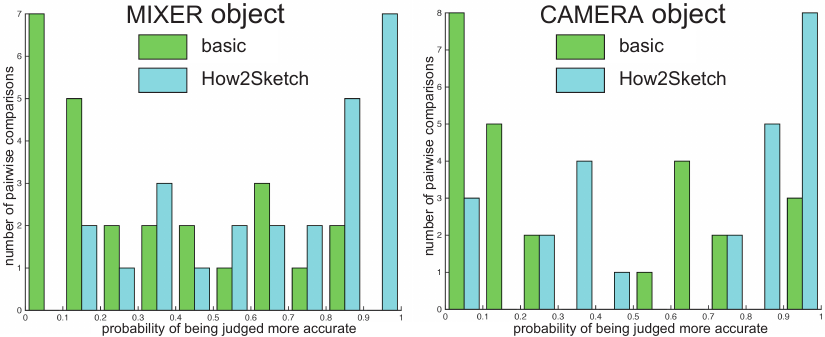}
\end{overpic}
\caption{Bradley-Terry Model for the Mixer and the Camera Sketches produced by users of our tutorials and evaluated by another user study with Amazon Mechanical Turk rankers.}
\label{fig:mixer_bt_plot}
\end{figure}

%
%
\mypara{Feedback} Nine out of ten participants preferred the How2Sketch tutorial over the Basic tutorial. User responses to questions about satisfaction with their drawing, perceived accuracy of their drawing, enjoyment of the drawing process, and ease of following tutorials were all higher for How2Sketch than the Basic tutorial~(Fig.~\ref{fig:survey}). 

An ANOVA across tutorial type and object drawn reveals a strong significant effect of tutorial type on accuracy and ease of following tutorial (p$<$0.003), significant effect on enjoyment (p$<$0.034), and marginally significant effect on satisfaction (p$<$0.058). The object drawn did not have an effect on any measure, despite their varying difficulty, and there was no interaction between tutorial type and object drawn. 

Freeform feedback echoed the ratings. Participants enjoyed using the guides as it gave them more confidence in their accuracy. One participant said ``It was satisfying drawing the guides and getting the proportions right. It was then so much easier to draw the final sketch using the blocks (scaffold) for guidance''. Users also had suggestions for improvements to the system with several asking for further guidance with perspective.

\mypara{Sketch quality} Figure \ref{fig:sketch_overlay} overlays the registered user sketches from the different conditions on the original model for the condition (e.g How2Sketch model after part level adjustments). While there appears to be variation in contour placement in both tutorial types the variation in the basic tutorial sketches is greater. 

In the camera tutorials the basic version starts with the ground plane for the lens and How2Sketch with the ground plane of the main body. With this anchoring in the basic tutorial sketches, the width and length of the lens are accurate. However, the lens height and the other three primitives have a variety of errors in proportion and part placement. Comparing with How2Sketch sketches, there are similar variations in the height of the main body. However, the guided steps for the grip and lens shows decent convergence in positioning across users. For the Mixer sketches - where both tutorials start with the base plane of the mixer - there is much more consistency of object part placements across users. 

To further validate these findings, we conducted an additional user study using Amazon Mechanical Turk (AMT). In the study, we presented users with two sketches of the same object type overlaid and registered to their condition specific model (see supplementary material). The two pairs of object could be from the same or different tutorial types. We asked participants to ``Please select the sketch that is more accurate to the underlying model. Do consider the proportions, alignment and perspective. Please ignore style or shading''. The studies for the two objects were run independently, so did not have the same responders. Each participant evaluated the 45 unique image pairs for one of the objects.  Each study had 225 sets of responses. AMT users we compensated \$0.01 per comparison. 

To evaluate the results, we performed pairwise comparisons using the Bradley-Terry model. We plot probability histograms for both objects in Figure~\ref{fig:mixer_bt_plot}. The bottom axis is the probability of being judged more accurate in a pairwise comparison. The left axis is the number of pairwise comparisons that has this probability. Visually you can see the trend that in a pairwise comparison sketches from following How2Sketch tutorials are more likely to be judged more accurate. This evidence is clear across both object types.

\section{Conclusion}

We presented How2Sketch, a system that automatically generates {\em easy-to-follow} tutorials for drawing  man-made 3D objects from selected views. 
%
%
%
We evaluated our system using a user study, and found that sketches made by following How2Sketch tutorials had more accurate proportions and relative part placements compared to a basic step-by-step tutorial with scaffolding primitives. Additionally, users preferred the How2Sketch tutorials over the basic tutorial, giving significantly higher ratings for satisfaction, accuracy, and enjoyment. 

One possible future direction is to provide stroke level support to help users draw the final object contours, possibly by explicitly providing guidelines with respect to the scaffold primitives. Another direction would to to explore new types of guidelines that can help reduce the number of unguided steps in a tutorial. 

A very interesting future question is to investigate if How2Sketch really {\em teaches} users to sketch better by drawing ``what you know." While this is the ultimate goal of any sketching tutorial, 
answering this question will require a much more involved user study where we have to track and quantify user-specific improvements, if any, as observed over a substantial timeframe. 

%
%

\bibliographystyle{acmsiggraph}
\bibliography{how2sketch}
\end{document}